\documentclass[11pt,a4paper]{article}

\usepackage[utf8]{inputenc}
\usepackage[T1]{fontenc}
\usepackage{cmbright}
\usepackage{microtype}
\usepackage[margin=2.5cm]{geometry}
\usepackage{graphicx}
\usepackage[round,sort]{natbib}
\usepackage[dvipsnames]{xcolor}
\usepackage{hyperref}
\hypersetup{breaklinks = true,
	        colorlinks = true,
	        linkcolor  = blue,
            urlcolor = RoyalPurple,
	        citecolor  = Blue}

\title{\vspace*{-2em}COSIMA Cookbook: a community-driven resource for ocean and sea-ice modelling science}

\author{%
\begin{minipage}{\dimexpr\textwidth-2\tabcolsep\relax}
\centering\normalsize
\href{https://orcid.org/0000-0002-8149-4094}{Navid~C.~Constantinou}\textsuperscript{3,14,*},
\href{https://orcid.org/0000-0002-3573-996X}{Julia~Neme}\textsuperscript{1,*},
\href{https://orcid.org/0000-0003-2453-9791}{Wilton~Aguiar}\textsuperscript{1},
\href{https://orcid.org/0000-0001-6228-5732}{Matthis~Auger}\textsuperscript{13},
\href{https://orcid.org/0000-0003-3165-8676}{Ashley~J.~Barnes}\textsuperscript{3,10},
\href{https://orcid.org/0000-0003-3891-5444}{Romain~Beucher}\textsuperscript{1,2},
\href{https://orcid.org/0000-0002-1222-375X}{Dhruv~Bhagtani}\textsuperscript{11},
\href{https://orcid.org/0000-0001-8362-3446}{Christopher~Yit~Sen~Bull}\textsuperscript{1,2},
\href{https://orcid.org/0000-0001-9113-1329}{Hannah~Dawson}\textsuperscript{17},
\href{https://orcid.org/0000-0003-4176-7956}{Noah~Day}\textsuperscript{14},
\href{https://orcid.org/0000-0002-2965-2120}{Fabio~Boeira~Dias}\textsuperscript{17},
\href{https://orcid.org/0000-0002-6097-5729}{Edward~W.~Doddridge}\textsuperscript{17},
\href{https://orcid.org/0009-0002-8898-0068}{Anupiya~Ellepola}\textsuperscript{1},
\href{https://orcid.org/0000-0002-5039-6272}{Denisse~Fierro-Arcos}\textsuperscript{17},
\href{https://orcid.org/0000-0001-7577-3604}{Angus~Gibson}\textsuperscript{1},
\href{https://orcid.org/0000-0002-4481-4896}{Aidan~Heerdegen}\textsuperscript{1,2},
\href{https://orcid.org/0000-0001-5898-7635}{Andy~McC.~Hogg}\textsuperscript{1,2,3},
\href{https://orcid.org/0000-0002-6799-9109}{Ryan~M.~Holmes}\textsuperscript{4},
\href{https://orcid.org/0000-0002-1476-0452}{Maurice~F.~Huguenin}\textsuperscript{8},
\href{https://orcid.org/0000-0001-8624-365X}{Wilma~Huneke}\textsuperscript{10},
\href{https://orcid.org/0000-0001-7190-7329}{Jemma~Jeffree}\textsuperscript{1,3},
\href{https://orcid.org/0000-0001-8960-9557}{Andrew~E.~Kiss}\textsuperscript{1},
\href{https://orcid.org/0000-0002-6167-9999}{Minghang~Li}\textsuperscript{1,2},
\href{https://orcid.org/0000-0003-3538-633X}{Paige~Martin}\textsuperscript{1,2},
\href{https://orcid.org/0000-0002-8348-1588}{Josué~Martínez-Moreno}\textsuperscript{12},
\href{https://orcid.org/0000-0001-8667-488X}{Jan~Jaap~Meijer}\textsuperscript{17},
\href{https://orcid.org/0000-0003-3930-1946}{Thomas~Moore}\textsuperscript{6},
\href{https://orcid.org/0000-0001-5054-1559}{Ruth~Moorman}\textsuperscript{5},
\href{https://orcid.org/0009-0003-5143-5020}{Adele~K.~Morrison}\textsuperscript{1},
\href{https://orcid.org/0000-0001-9098-6309}{James~Munroe}\textsuperscript{7},
\href{https://orcid.org/0000-0002-8967-2211}{Aditya~Narayanan}\textsuperscript{16},
\href{https://orcid.org/0000-0003-1364-0907}{Micael~Oliveira}\textsuperscript{1,2},
\href{https://orcid.org/0000-0002-0392-7915}{Ellie~Q.~Y.~Ong}\textsuperscript{3,10},
\href{https://orcid.org/0000-0003-4254-843X}{Madelaine~G.~Rosevear}\textsuperscript{3,17},
\href{https://orcid.org/0000-0002-7672-5054}{Christina~Schmidt}\textsuperscript{12},
\href{https://orcid.org/0000-0002-4162-3269}{Taimoor~Sohail}\textsuperscript{14},
\href{https://orcid.org/0000-0001-5156-2204}{Paul~Spence}\textsuperscript{3,17},
\href{https://orcid.org/0000-0003-3271-6874}{Dougal~T.~Squire}\textsuperscript{1,2},
\href{https://orcid.org/0009-0002-9081-4106}{Anton~Steketee}\textsuperscript{1,2},
\href{https://orcid.org/0000-0002-3262-4972}{Charles~Turner}\textsuperscript{1,2},
\href{https://orcid.org/0000-0003-1967-880X}{Felipe~Vilela~da~Silva}\textsuperscript{9},
\href{https://orcid.org/0000-0003-3882-418X}{Marc~White}\textsuperscript{1,2},
\href{https://orcid.org/0000-0001-8570-7424}{Luwei~Yang}\textsuperscript{1},
\href{https://orcid.org/0000-0002-0052-7668}{Claire~Yung}\textsuperscript{1},
\href{https://orcid.org/0000-0003-3462-3559}{Jan~Zika}\textsuperscript{15}
\\[2ex]
\footnotesize
\textsuperscript{1}Australian National University, Australia\\
\textsuperscript{2}Australia's Climate Simulator (ACCESS-NRI), Australia\\
\textsuperscript{3}Australian Research Council Centre of Excellence for the Weather of the 21st Century, Australia\\
\textsuperscript{4}Bureau of Meteorology, Australia\\
\textsuperscript{5}California Institute of Technology, USA\\
\textsuperscript{6}Commonwealth Scientific and Industrial Research Organisation, Australia\\
\textsuperscript{7}Earthdata Associates, Canada\\
\textsuperscript{8}GEOMAR Helmholtz Centre for Ocean Research Kiel, Germany\\
\textsuperscript{9}IFREMER - French Institute for Ocean Science, France\\
\textsuperscript{10}Monash University, Australia\\
\textsuperscript{11}Princeton University, USA\\
\textsuperscript{12}British Antarctic Survey, UK\\
\textsuperscript{13}University of Brest, France\\
\textsuperscript{14}University of Melbourne, Australia\\
\textsuperscript{15}University of New South Wales Sydney, Australia\\
\textsuperscript{16}University of Southampton, UK\\
\textsuperscript{17}University of Tasmania, Australia\\
\textsuperscript{*}These authors contributed equally.\\[1em]
Software: \href{https://github.com/COSIMA/cosima-recipes}{github.com/COSIMA/cosima-recipes}\\
Zenodo: \href{https://doi.org/10.5281/zenodo.14353852}{https://doi.org/10.5281/zenodo.14353852}
\end{minipage}
}

\date{}

\begin{document}

\maketitle

\vspace{-2em}
\begin{abstract}
The COSIMA Cookbook is an open educational resource for analysing ocean and sea-ice model output in Jupyter notebooks. It has been developed by the Consortium for Ocean--Sea Ice Modelling in Australia (COSIMA) as a community resource for researchers, students, and practitioners working with large gridded datasets, especially output from the ACCESS ocean--sea ice model configurations. The Cookbook is organised around a gastronomy theme: ``Cooking Tutorials'' teach generic, transferable techniques for handling model output, while ``Easy Recipes'', ``Advanced Recipes'' and ``Regional Specialties'' are self-contained notebooks that show how to perform concrete diagnostics and analyses. Along the way, the notebooks demonstrate open-source tools for the scientific analysis of large data, including Intake catalogues, Xarray, Dask, xgcm, xESMF, eofs, Parcels, matplotlib and Cartopy. The Cookbook addresses the steep entry cost of analysing climate model output: it packages expert, often tacit, knowledge into complete, well-documented and reusable workflows, in the same medium in which researchers actually work. It is developed openly on GitHub, is published as a documentation website, and is continuously reviewed and extended by its community through pull requests and regular hackathons. The easy access to data and analysis it provides has contributed to the international adoption of ACCESS ocean and sea-ice model configurations, and has facilitated well over 100 peer-reviewed papers and more than 20 PhD projects.
\end{abstract}

\section{Summary}

\begin{figure}[htbp]
  \centering
  \includegraphics[width=0.85\textwidth]{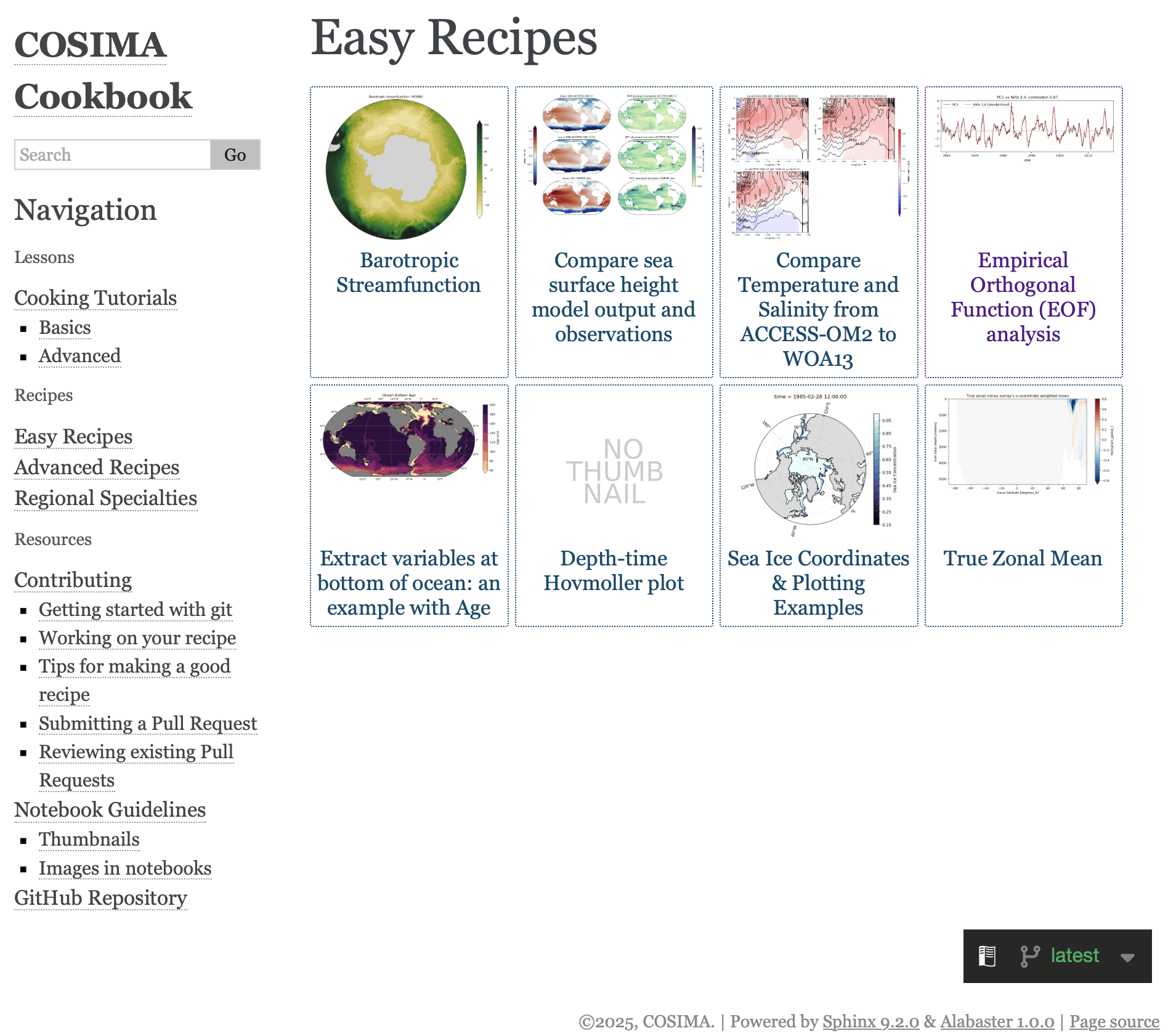}
  \caption{The COSIMA Cookbook documentation website, showing the browsable Sphinx-based landing page used to navigate tutorials and recipes. The website lives at \url{https://cosima-recipes.readthedocs.io}.}
  \label{fig:website}
\end{figure}

The COSIMA Cookbook is an open educational resource for analysing
ocean and sea-ice model output in Jupyter notebooks \citep{kluyver2016jupyter}.
It has been developed by the Consortium for Ocean--Sea Ice Modelling in Australia
(COSIMA; \url{https://cosima.org.au}) as a community resource for researchers,
students, and practitioners working with large gridded datasets, especially output
from the ACCESS ocean--sea ice model configurations \citep[e.g.][]{kiss2020accessom2}.
The Cookbook is developed openly in a GitHub repository, with all content exposed
through a documentation site built with Sphinx \citep{sphinx}; see \autoref{fig:website}.
Tagged versions of the repository are uploaded to a citable archived release \citep{cosimaCookbook}.

The Cookbook grew out of a loosely organised collection of Python scripts and Jupyter
notebooks---many of which were broken. During the first COVID-19 lockdown in 2020,
the community held its first hackathon \citep{cosima_hackathon_2020}: a full-day online sprint
to bring those notebooks up to date. Annual workshops and hackathons since then have
gradually shaped the repository into its current curated state.

The COSIMA Cookbook is deliberately organized using a gastronomy theme to help users
quickly understand and navigate the website. ``Cooking Tutorials'' refers to
tutorials that teach generic, transferable techniques for handling ocean--sea ice
model output. These tutorials give the users the foundations for understanding the analysis
done in the examples. Then comes the main part of any cookbook -- its recipes.
Here, by ``recipes'' we mean self-contained notebooks showing how to perform concrete
diagnostics and analyses on ocean and sea-ice datasets. First come the ``Easy Recipes'' that
provide a good starting point after the tutorials; then the ``Advanced Recipes''
that are more elaborate and demonstrate advanced analysis techniques. Lastly, the
``Regional Specialties'' contains recipes for regional model configurations \citep{barnes2024regionalmom6}.
This gastronomy framing is shared by other computational workflow collections, such as
\href{https://cookbooks.projectpythia.org}{Project Pythia Cookbooks} for the general
geoscience community and the \href{https://faircookbook.elixir-europe.org/content/home.html}{FAIR Cookbook}
for working with life science data.

Pedagogy is central to the project. ``Cooking Tutorials'' introduces generic skills such
as loading model output, working with labelled arrays, plotting, and interacting with
shared data catalogues. These tutorials lead into domain-focused recipes giving learners an incremental path from first contact with the data ecosystem to adaptation of full
workflows for their own science questions. The recipes provide the starting point for
more elaborate analyses. The categories of ``easy'' and ``advanced'' recipes provide
a lightweight pedagogical cue about expected complexity and scope.

Within this structure, the tutorials leverage and demonstrate open source tools for
scientific analysis of large data. This includes examples of loading model output
through Intake-based catalogues \citep{intake}, using Xarray \citep{hoyer2017xarray}
and Dask \citep{rocklin2015dask,dask} to process very large multi-dimensional datasets
in parallel. The recipe collection includes a Lagrangian particle-tracking workflow
built with Parcels \citep{Delandmeter-vanSebille-2019}, as well as analyses that use
xgcm \citep{xgcm} to handle variables defined on staggered finite-volume grids, xESMF \citep{xesmf}
for regridding model output, and eofs \citep{dawson2016eofs} for empirical orthogonal function analysis. Figures are generated using matplotlib \citep{Hunter2007matplotlib}
incorporating Cartopy \citep{cartopy} for map-based visualisation.

The Cookbook grew from a practical need inside the COSIMA community \citep{cosima_movie}:
powerful tools for analysing modern ocean model output already existed, but the gap between
package-level documentation and reproducible end-to-end workflows left new users
largely on their own. Productive use requires fluency not only in Python and Jupyter
\citep{kluyver2016jupyter}, but also in navigating high-dimensional, often very large model output,
shared high-performance computing environments, and domain-specific analysis
conventions. The Cookbook bridges this gap by packaging reusable workflows in
the same medium in which researchers actually work.

\section{Statement of Need}

Analysing or configuring climate models carries a steep entry cost. New users must learn
how to find datasets, load them efficiently, interpret metadata,
operate on multi-dimensional arrays, and produce scientifically meaningful diagnostics.
General-purpose libraries such as Xarray \citep{hoyer2017xarray} provide the
foundations for these tasks, but they do not by themselves show a learner how
to translate a research question into a robust analysis workflow for a specific
model and computing environment. The Cookbook serves as a learning tool
for those new to ocean and sea ice science, showing how to interrogate model output
and visualise commonly used diagnostics---skills that transfer broadly across models
and computing environments.

The gap is not only pedagogical; it directly slows science. Before any analysis can begin,
a newcomer typically spends substantial time deciphering the structure of the output,
the relevant conventions, and the computational environment. There is also
substantial scientific labour embedded in many common diagnostics:
some calculations require non-trivial development, understanding
of model numerics, and validation before they can be used with confidence---these are the
``Advanced Recipes''. If every new user had to reconstruct those workflows independently,
enormous effort would be spent rebuilding analysis code rather than doing science.

The COSIMA Cookbook facilitates knowledge sharing and accelerates research with
a domain-specific, openly maintained collection of computational lessons and examples.
Its contribution is not a new analysis library; rather, it is a reusable scientific
and educational resource that exposes complete, well-documented workflows in the same
notebook format in which many researchers actually explore data and communicate methods.
By choosing notebooks instead of packaging these examples only as a software
library, the project makes intermediate reasoning, methodological choices, and
practical execution details visible to learners while still giving more
experienced users analysis patterns they can adapt directly for research. This
aligns well with JOSE's emphasis on open educational resources that enable
computational learning through authentic practice.

Several design decisions have made the Cookbook particularly useful for adoption by
other groups. First, each notebook is intended to be self-contained and
well-documented, so learners can read, run, and modify a complete workflow
without reconstructing missing context from scattered notes. Second, the
repository distinguishes tutorials from recipes. Tutorials teach transferable
skills, while recipes demonstrate how those skills combine in realistic
analysis tasks. Third, the project includes contributor guidance and notebook
review conventions that encourage new analyses to be written in a reusable and
pedagogically clear style. A caveat is that most recipes use high-resolution
model output with datasets of order terabytes, so the underlying data are not
always practical to distribute or mirror in full. However, beyond the data-loading step,
which usually happens early on in each recipe, the analysis workflows are generally
applicable to most operational ocean--sea ice model outputs that can be
loaded with Xarray \citep{hoyer2017xarray}.

Thus, the design renders the Cookbook valuable beyond the immediate COSIMA context.
Many research communities maintain model output on shared infrastructure and face the same
challenge: turning expert tacit knowledge into examples that newcomers can
adapt. The Cookbook offers a model for how a scientific collaboration can
capture that knowledge in version-controlled notebooks, publish it as a living open
resource online, and continuously improve it through community contribution.

The rise of AI-assisted coding tools adds a further dimension to the Cookbook's value---and arguably makes it more important, not less. Such tools can rapidly generate analysis
code, and they produce far better results when given a well-structured, domain-specific
recipe as context rather than starting from scratch. But the deeper point is this: as
AI-generated code becomes routine, the critical skill shifts from \emph{writing} code to
\emph{understanding and evaluating} it. A researcher who has worked through the Cookbook knows
why a workflow is structured the way it is---which parameters matter, where scientific
assumptions enter, and what a plausible result looks like. That understanding is what
allows AI-generated code to be trusted, corrected, and adapted for specific science
questions. Resources that teach the reasoning behind an analysis, not just its mechanics,
therefore grow more valuable alongside AI adoption.

\section{Educational Design and Experience of Use}

As we have elaborated above, the learning experience is organised as a progression.
The tutorials and recipes showcase Jupyter-based analysis that can be run directly
on Australia's National Computational Infrastructure, and include guidance on
running notebooks through JupyterLab sessions and on accessing shared data holdings.
This makes the Cookbook more than a static collection of examples: it is operational
documentation for a real analysis environment. At the same time, the notebooks
demonstrate broadly transferable workflows for working with labelled geophysical data,
so individual lessons can be reused or adapted for specific research questions or
for use on other computational platforms.

The project also supports community authorship. Contributors are encouraged to
submit new notebooks through pull requests, document their methods clearly, and
generalise workflows so that they remain useful to others. This is particularly valuable as
it is relatively rare for the computer code underlying scientific analyses to be peer reviewed
with the same care as the manuscript itself. The COSIMA community provides several examples of
the recipe ecosystem being extended into full research projects built on peer-authored software
repositories on GitHub, where analysis workflows, methods, and implementation choices
are exposed to community scrutiny and reuse. In that sense, the Cookbook functions both
as a learning resource for end users and as a framework for teaching reproducible scientific
communication through notebook design and peer review. During regularly organised hackathons,
the community reviews, updates, and extends existing recipes, ensuring the Cookbook remains
a living resource that keeps pace with evolving tools and community best practices.
The easy access to data and analysis provided by the COSIMA Cookbook has enabled rapid
adoption of ACCESS ocean and sea-ice model configurations internationally---not only within
physical oceanography but also across adjacent disciplines such as marine ecology
\citep{FierroArcos2023}, where researchers have used it as an entry point to ACCESS model output.
This accessibility has facilitated well over
\href{https://scholar.google.com/citations?hl=en&user=inVqu_4AAAAJ}{100 peer-reviewed papers}
and more than 20 PhD projects to completion to date.

\section*{Author Order}

The first two authors contributed equally; authors from the third position
onward are listed alphabetically by surname.

\section*{Acknowledgements}

This work was supported by computational resources provided by the Australian Government
through the National Computational Infrastructure (NCI) under the National Computational Merit
Allocation Scheme and the Australian National University Allocation Scheme. We thank the
vibrant communities of the Consortium for Ocean--Sea Ice Modelling in Australia (COSIMA;
\href{https://cosima.org.au}{cosima.org.au}) and Australia's climate simulator (ACCESS-NRI;
\href{https://access-nri.org.au}{access-nri.org.au}) for making the ACCESS-OM2 and ACCESS-OM3 outputs
and analysis tools available through the NCI.
We further acknowledge funding from the Australian Research Council under the Centres of
Excellence for Climate Extremes CE170100023 and for the Weather of the 21st Century CE230100012, the Linkage Infrastructure, Equipment and Facilities LP160100073 and LP200100406, and the Discovery Project DP240101274.

\bibliographystyle{plainnat}
\bibliography{references}

@misc{cosima_hackathon_2020,
  author       = {Mart{\'i}nez-Moreno, Josu{\'e} and Constantinou, Navid C.},
  title        = {{Promotional video for COSIMA Hackathon April 2020}},
  year         = {2020},
  howpublished = {\url{https://youtu.be/JO-_jaPW8S8}},
  note         = {YouTube video}
}

@misc{cosima_movie,
  author       = {{COSIMA Community}},
  title        = {{COSIMA: The Movie}},
  year         = {2026},
  howpublished = {\url{https://youtu.be/bmUYRoPKtIA}},
  note         = {YouTube video}
}

@misc{cosimaCookbook,
  title = {{COSIMA Cookbook: A cookbook of recipes for analysing ocean and sea ice model output}},
  author       = {Constantinou, Navid C. and
                  Hogg, Andy McC. and
                  Gibson, Angus and
                  Steketee, Anton and
                  Beucher, Romain and
                  Proft, Max and
                  Heerdegen, Aidan and
                  Neme, Julia and
                  Holmes, Ryan M. and
                  Morrison, Adele and
                  Fierro-Arcos, Denisse and
                  Kiss, Andrew E. and
                  Oliveira, Micael and
                  Yung, Claire and
                  Doddridge, Edward W. and
                  Huneke, Wilma and
                  Bhagtani, Dhruv and
                  Ong, Ellie Q. Y. and
                  Munroe, James and
                  Schmidt, Christina and
                  Dawson, Hannah and
                  Squire, Dougal T. and
                  Dias, Fabio Boeira and
                  Jeffree, Jemma and
                  Moorman, Ruth and
                  Martínez-Moreno, Josué and
                  Zika, Jan and
                  Meijer, Jan Jaap and
                  Auger, Matthis and
                  Aguiar, Wilton and
                  Moore, Thomas and
                  Vilela da Silva, Felipe and
                  Yang, Luwei and
                  White, Marc and
                  Sohail, Taimoor and
                  Barnes, Ashley J. and
                  Spence, Paul and
                  Turner, Charles and
                  Rosevear, Madelaine G. and
                  Bull, Christopher Yit Sen and
                  Curtin, Katja and
                  Day, Noah and
                  Li, Minghang and
                  Ellepola, Anupiya and
                  Narayanan, Aditya and
                  Huguenin, Maurice F. and
                  Martin, Paige and
                  Meng, Hangyu and
                  Sholeninova, Polina},
  year         = 2026,
  publisher    = {Zenodo},
  version      = {v0.1.8},
  doi          = {10.5281/zenodo.24457957},
  url          = {https://doi.org/10.5281/zenodo.24457957},
}

@misc{sphinx,
  author       = {Komiya, Takeshi and
                  Brandl, Georg and
                  Turner, Adam and
                  Bakosi, Jean-Francois and
                  Shimizukawa, Takayuki and
                  Andersen, Jakob Lykke and
                  Neuh{\"a}user, Daniel and
                  Finucane, Stephen and
                  Lehmann, Robert and
                  Mason, Jacob and
                  Kampik, Timotheus and
                  Addison, James and
                  Magin, Justus and
                  Dufresne, Jon and
                  Waltman, Jonathan and
                  Eades, Daniel and
                  Cano Rodr{\'i}guez, Juan Luis and
                  Ronacher, Armin and
                  Geier, Matthias and
                  Shachnev, Dmitry and
                  Shibukawa, Yoshiki and
                  Tran, B{\'e}n{\'e}dikt and
                  Dangoor, Adam and
                  Sewell, Chris and
                  Ruana, Rob and
                  Virtanen, Pauli and
                  Hoffmann, Tim and
                  Li{\v{s}}ka, Martin and
                  Freitag, Fran{\c c}ois},
  title        = {{sphinx-doc/sphinx: Sphinx 9.1.0}},
  year         = {2025},
  publisher    = {Zenodo},
  version      = {v9.1.0},
  doi          = {10.5281/zenodo.18109073},
  url          = {https://doi.org/10.5281/zenodo.18109073}
}

@article{kiss2020accessom2,
  author = {Kiss, Andrew E. and Hogg, Andrew McC. and Hannah, Nicholas and Boeira Dias, Fabio and Brassington, Gary B. and Chamberlain, Matthew A. and Chapman, Christopher and Dobrohotoff, Peter and Domingues, Catia M. and Duran, Earl R. and England, Matthew H. and Fiedler, Russell and Griffies, Stephen M. and Heerdegen, Aidan and Heil, Petra and Holmes, Ryan M. and Klocker, Andreas and Marsland, Simon J. and Morrison, Adele K. and Munroe, James and Nikurashin, Maxim and Oke, Peter R. and Pilo, Gabriela S. and Richet, Oc{\'e}ane and Savita, Abhishek and Spence, Paul and Stewart, Kial D. and Ward, Marshall L. and Wu, Fanghua and Zhang, Xihan},
  title = {{ACCESS-OM2 v1.0: a global ocean-sea ice model at three resolutions}},
  journal = {Geoscientific Model Development},
  year = {2020},
  volume = {13},
  number = {2},
  pages = {401--442},
  doi = {10.5194/gmd-13-401-2020},
  url = {https://doi.org/10.5194/gmd-13-401-2020}
}

@article{hoyer2017xarray,
  author = {Hoyer, Stephan and Hamman, Joseph J.},
  title = {{xarray: N-D labeled arrays and datasets in Python}},
  journal = {Journal of Open Research Software},
  year = {2017},
  volume = {5},
  number = {1},
  pages = {10},
  doi = {10.5334/jors.148},
  url = {https://doi.org/10.5334/jors.148}
}

@incollection{kluyver2016jupyter,
  author = {Kluyver, Thomas and Ragan-Kelley, Benjamin and P{\'e}rez, Fernando and Granger, Brian and Bussonnier, Matthias and Frederic, Jonathan and Kelley, Kyle and Hamrick, Jessica and Grout, Jason and Corlay, Sylvain and Ivanov, Paul and Avila, Dami{\'a}n and Abdalla, Safia and Willing, Carol},
  title = {Jupyter Notebooks---a publishing format for reproducible computational workflows},
  booktitle = {Positioning and Power in Academic Publishing: Players, Agents and Agendas},
  editor = {Loizides, Fernando and Schmidt, Birgit},
  publisher = {IOS Press},
  address = {Amsterdam},
  year = {2016},
  pages = {87--90},
  doi = {10.3233/978-1-61499-649-1-87},
  url = {https://doi.org/10.3233/978-1-61499-649-1-87}
}

@article{barnes2024regionalmom6,
  author = {Barnes, Ashley J. and Constantinou, Navid C. and Gibson, Angus H. and Kiss, Andrew E. and Chapman, Chris and Reilly, John and Bhagtani, Dhruv and Yang, Luwei},
  title = {{regional-mom6: A Python package for automatic generation of regional configurations for the Modular Ocean Model 6}},
  journal = {Journal of Open Source Software},
  year = {2024},
  volume = {9},
  number = {100},
  pages = {6857},
  doi = {10.21105/joss.06857},
  url = {https://doi.org/10.21105/joss.06857}
}

@misc{cartopy,
  author       = {Phil Elson and
                  Elliott Sales de Andrade and
                  Greg Lucas and
                  Ryan May and
                  Richard Hattersley and
                  Ed Campbell and
                  Ruth Comer and
                  Andrew Dawson and
                  Bill Little and
                  Stephane Raynaud and
                  scmc72 and
                  Alan D. Snow and
                  lgolston and
                  Byron Blay and
                  Peter Killick and
                  lbdreyer and
                  Patrick Peglar and
                  Nat Wilson and
                  Andrew and
                  Jon Szymaniak and
                  Adrien Berchet and
                  Corinne Bosley and
                  Luke Davis and
                  Filipe and
                  John Krasting and
                  Matthew Bradbury and
                  stephenworsley and
                  Daniel Kirkham},
  title        = {{SciTools/cartopy: v0.24.1}},
  month        = oct,
  year         = 2024,
  publisher    = {Zenodo},
  version      = {v0.24.1},
  doi          = {10.5281/zenodo.13905945},
  url          = {https://doi.org/10.5281/zenodo.13905945},
}

@manual{intake,
  author = {{Intake developers}},
  title = {Intake: a lightweight package for finding, investigating, loading and disseminating data},
  year = {2026},
  url = {https://intake.readthedocs.io/en/reader/}
}

@Article{Delandmeter-vanSebille-2019,
  author = {Delandmeter, Philippe and van Sebille, Erik},
  title = {The {Parcels} v2.0 {Lagrangian} framework: new field interpolation schemes},
  journal = {Geoscientific Model Development},
  volume = {12},
  year = {2019},
  number = {8},
  pages = {3571--3584},
  doi = {10.5194/gmd-12-3571-2019}
}

@manual{dask,
  title = {Dask: Library for dynamic task scheduling},
  author = {{Dask Development Team}},
  year = {2016},
  url = {http://dask.pydata.org},
}

@misc{xgcm,
  author       = {Abernathey, Ryan P. and
                  Busecke, Julius J. M. and
                  Smith, Timothy Andrew and
                  Deauna, Josephine Dianne and
                  Banihirwe, Anderson and
                  Nicholas, Thomas and
                  Fernandes, Filipe and
                  James, Bourbeau and
                  Dussin, Raphael and
                  Cherian, Deepak A. and
                  Caneill, Romain and
                  Sinha, Anirban and
                  Uieda, Leonardo and
                  Rath, Willi and
                  Balwada, Dhruv and
                  Constantinou, Navid C. and
                  Ponte, Aur{\'e}lien and
                  Zhou, Yuxin and
                  Uchida, Takaya and
                  Thielen, Jon},
  title        = {{xgcm}},
  month        = nov,
  year         = 2022,
  publisher    = {Zenodo},
  version      = {v0.8.1},
  doi          = {10.5281/zenodo.7348619},
  url          = {https://doi.org/10.5281/zenodo.7348619},
}

@article{dawson2016eofs,
  author       = {Andrew Dawson},
  title        = {eofs: A Library for {EOF} Analysis of Meteorological, Oceanographic, and Climate Data},
  journal      = {Journal of Open Research Software},
  year         = {2016},
  volume       = {4},
  number       = {1},
  pages        = {e14},
  doi          = {10.5334/jors.122},
}

@misc{xesmf,
  author       = {Jiawei Zhuang and
                  Raphael Dussin and
                  Pascal Bourgault and
                  David Huard and
                  Anderson Banihirwe and
                  Stephane Raynaud and
                  Brewster Malevich and
                  Martin Schupfner and
                  Charles Gauthier and
                  Filipe and
                  Sam Levang and
                  Mattia Almansi and
                  Andr{\'e} J{\"u}ling and
                  Richard Scott and
                  RondeauG and
                  Stephan Rasp and
                  Trevor James Smith and
                  Aaron G. Meyer and
                  Ben Mares and
                  Bill Sacks and
                  Jemma Stachelek and
                  Matthew Plough and
                  Pierre and
                  Ray Bell and
                  Romain Caneill and
                  Xianxiang Li},
  title = {{pangeo-data/xESMF: v0.9.2}},
  year = {2025},
  doi = {10.5281/zenodo.4294774},
  url = {https://doi.org/10.5281/zenodo.4294774}
}

@InProceedings{rocklin2015dask,
  author    = {Matthew Rocklin},
  title     = {Dask: Parallel Computation with Blocked algorithms and Task Scheduling},
  booktitle = {Proceedings of the 14th Python in Science Conference},
  pages     = {130--136},
  year      = {2015},
  editor    = {Kathryn Huff and James Bergstra},
  doi       = {10.25080/Majora-7b98e3ed-013}
}

@Article{Hunter2007matplotlib,
  Author    = {Hunter, John D.},
  Title     = {Matplotlib: A {2D} graphics environment},
  Journal   = {Computing in Science \& Engineering},
  Volume    = {9},
  Number    = {3},
  Pages     = {90--95},
  publisher = {IEEE COMPUTER SOC},
  doi       = {10.1109/MCSE.2007.55},
  year      = 2007
}

@article{FierroArcos2023,
  title = {Analysis of ecologically relevant sea ice and ocean variables for the {Southern Ocean} using a high-resolution model to inform ecosystem studies},
  journal = {Progress in Oceanography},
  volume = {215},
  pages = {103049},
  year = {2023},
  doi = {10.1016/j.pocean.2023.103049},
  author = {Denisse Fierro-Arcos and Stuart Corney and Amelie Meyer and Hakase Hayashida and Andrew E. Kiss and Petra Heil},
}

\end{document}